\documentclass[twoside]{dis08}
\usepackage[latin1]{inputenc}
\usepackage[dvips]{graphicx,epsfig,color}
\usepackage{wrapfig,rotating}
\usepackage{amssymb,amsmath,array}

\newcommand{\beq}{\begin{equation}}
\newcommand{\eeq}{\end{equation}}
\newcommand{\beqy}{\begin{eqnarray}}
\newcommand{\eeqy}{\end{eqnarray}}
\newcommand{\beqyn}{\begin{eqnarray*}}
\newcommand{\eeqyn}{\end{eqnarray*}}

\begin{document}
\title{Gluon Polarization and Higher Twist Effects}

%***********************************************************************
% AUTHORS INFORMATION AREA
%***********************************************************************
\author{Elliot Leader$^1$ , Alexander Sidorov$^2$\thanks{Supported by RFBR
Grants 06-02-16215 and 08-01-00686} ~and Dimiter Stamenov$^3$
%
% Optional short acknowledgment: remove next line if non-needed
%
% DO NOT MODIFY THE FOLLOWING '\vspace' ARGUMENT
\vspace{.3cm}\\
%
% Addresses and institutions (remove "1- " in case of a single institution)
1- Imperial College London \\
London - UK
%
% Remove the next three lines in case of a single institution
\vspace{.1cm}\\
2- Bogolubov Theoretical Laboratory, JINR \\
Dubna - Russia
\vspace{.1cm}\\
3- Institute for Nuclear Research and Nuclear Energy \\
Bulgarian Academy of Sciences, Sofia - Bulgaria }
%***********************************************************************
% END OF AUTHORS INFORMATION AREA
%***********************************************************************

\maketitle

\begin{abstract}
We examine the influence of the recent CLAS and COMPASS
experiments on our understanding of higher twist (HT) effects and
the gluon polarization, and show how EIC could discriminate
between negative and positive gluon polarizations. We comment on
the issue of HT and the recent DSSV analysis.
\end{abstract}

\section{Higher Twist (HT) Effects}

%\vspace{-0.4cm}
\begin{wrapfigure}{r}{0.47\columnwidth}
\vspace{-1.4cm}
\center{\includegraphics[width=0.46\columnwidth]{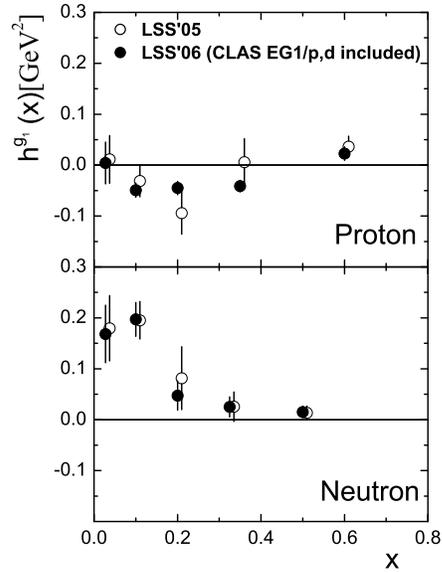}}
\caption{Higher twist terms for protons and
neutrons.}\label{Fig:1} \label{delGovG}
\end{wrapfigure}
CLAS ~\cite{1} has presented very accurate data on $g_1^p$ and
$g_1^d$ at low $Q^2 \, (1-4\, GeV^2)$ and $0.1 \leq x \leq0.6$,
yielding an improvement in the determination of  HT effects in
$g_1(x)$. Compared to the HT values obtained in the LSS'05
analysis \cite{2}, the uncertainties in the HT values at each $x$
are significantly reduced by the CLAS data, as  seen in Fig.
\ref{Fig:1}. (For details see \cite{3}, where $7$ $x$-bins were
used. Results in the present paper are based on an extraction of
the HT terms in $5$ $x$-bins. )
%\begin{figure}[h]
%\center{\includegraphics[width=0.45\columnwidth]{HT_CLAS2.eps}}
%\caption{Higher twist terms for protons and
%neutrons.}\label{Fig:1}
%\end{figure}

%%%%%%%%%%%%%%%%%%%%%%%%%%%%%%%%%%%%%%%%%%%%%%%%%%%%%%%%%%%%%5

Long ago we observed empirically that we could fit the the ratio
$\frac{g_1}{F_1}$ without any higher twist terms. If we split
$g_1$ and $F_1$ into leading and higher twist pieces

\beq g_1 = g_1^{LT} + g_1^{HT} \qquad  F_1 = F_1^{LT} + F_1^{HT}
\nonumber \eeq

then, approximately,

\beq \frac{g_1}{F_1}\approx \frac{g_1^{LT}}{F_1^{LT}}\, \big [ 1 +
\frac{g_1^{HT}}{g_1^{LT}} - \frac{F_1^{HT}}{F_1^{LT}}\big ]
\nonumber \eeq

%\newpage

\begin{wrapfigure}{r}{0.50\columnwidth}
\centerline{\epsfysize=2.2in
\epsfxsize=2.4in\epsfbox{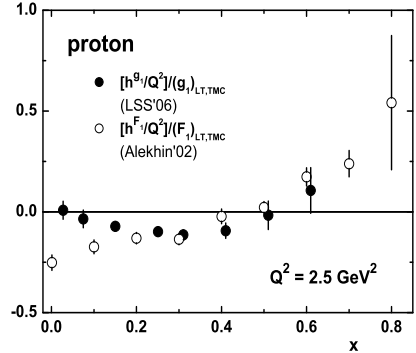}}
%\center{\includegraphics[width=0.52\columnwidth]{Al02_g1HTovg1_F1HTovF1.EPS}}
\caption{Comparison of HT terms in $g_1$ and $F_1$.}\label{Fig:2}
\end{wrapfigure}

Thus our observation requires a cancellation between
${g_1^{HT}}/{g_1^{LT}}$ and ${F_1^{HT}}/{F_1^{LT}}$. Fig.
\ref{Fig:2}, based on our recent results on $g_1$ ~\cite{3} and
the unpolarized results of ~\cite{Alekhin}, demonstrates the
validity of this for $x\geq 0.15$, but clearly indicates that
ignoring HT terms in the ratio $ \frac{g_1}{F_1}$ below $x=0.15$,
as was done in the recent DSSV analysis ~\cite{deFlorian:2008mr},
is incorrect.

%\vspace*{-1.4cm}

%\begin{figure}[h]
%\center{\includegraphics[width=0.45\columnwidth]{HTratiosg1F1.eps}}
%\caption{Comparison of HT terms in $g_1$ and $F_1$.}\label{Fig:A}
%\end{figure}

%%%%%%%%%%%%%%%%%%%%%%%%%%%%%%%%%%%%%%%%%%%%%%%%%%%%%%%%%%%%%%
%\vspace{-1.4cm}

 COMPASS ~\cite{4} has  presented data on $g_1^d(x)$ at
 large $Q^2$ and very small $x \, (0.004\leq x \leq 0.02)$,
 the \emph{only} precise data at such small $x$. Their influence on
 the  HT terms is negligible (see the talk of Sidorov at the XII Workshop
 on High Energy Spin Physics, Dubna, 2007  ~\cite{5}), but they significantly  effect
  the extraction of the polarized gluon density.
%\smallskip

\section{The polarized gluon density}

\begin{wrapfigure}{r}{0.50\columnwidth}
\includegraphics[width=0.49\columnwidth]{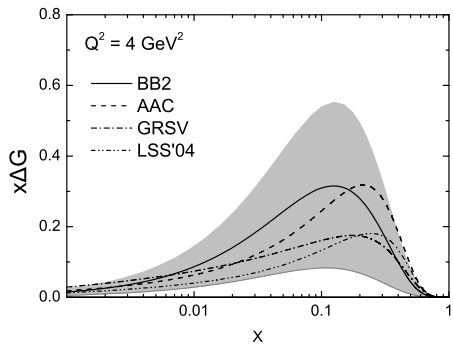}
\caption{Results of various analyses for $\Delta G$.}\label{Fig:3}
\end{wrapfigure}

There are three ways to access $\Delta G(x)$:

%\begin{figure}[h]
%\includegraphics[width=0.45\columnwidth]{VariousDelG_BW.eps}
%\caption{Results of various analyses for $\Delta G$.}\label{Fig:2}
%\end{figure}
%\begin{wrapfigure}{r}{0.6\columnwidth}
%\includegraphics[width=0.45\columnwidth]{VariousDelG_BW.eps}
%\caption{Results of various analyses for $\Delta G$.}\label{Fig:2}
%\end{wrapfigure}
(i) via polarized DIS: we parametrize the polarized quark and
gluon densities and fit data on $g_1(x, Q^2)$. The main role of
the gluon is in the \emph{evolution} with $Q^2$, but the range of
$Q^2$ is very
 limited, so the determination of $\Delta G(x)$ is
imprecise.
 For a long time various analyses seemed to indicate
that $\Delta G(x)$ was a positive function of $x$, as seen in
Fig.~\ref{Fig:3} .

%\begin{figure}[h]
%\begin{center}
%\includegraphics[width=0.45\columnwidth]{VariousDelG_BW.eps}
%\caption{Results of various analyses for $\Delta G$.}\label{Fig:2}
%\end{center}
%\end{figure}

For reasons that we do not understand, with the inclusion of
recent data, we get equally good fits with positive, negative and
sign-changing $\Delta G(x)$, provided we include higher twist
terms. The latter are particularly demanded by the CLAS data. Note
that the COMPASS analysis finds acceptable negative $\Delta G(x)$
fits, but has some peculiarities, which suggest it is not very
physical. They do \emph{not} include HT terms! We fail to find
negative $\Delta G(x)$ fits \emph{without} HT terms!

\begin{figure}[tbh]
\begin{center}
\includegraphics[width=0.44\columnwidth]{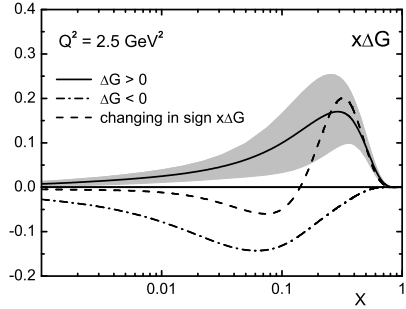}
\caption{The three different LSS'06 parametrizations of $\Delta G$
.} \label{Fig:4}
\end{center}
\end{figure}

\begin{figure}[tbh]
\begin{minipage}[c]{6.5cm}
\includegraphics[width = 6cm]{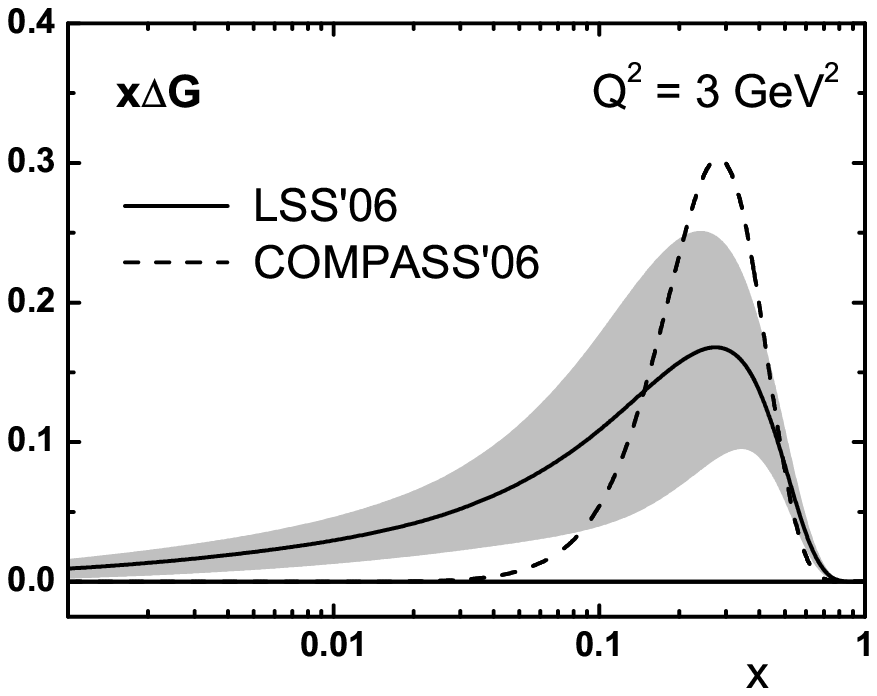}
\end{minipage}
\hspace*{0.5cm}
\begin{minipage}[c]{6.5cm}
\vspace*{-0.35cm}
\includegraphics[width= 6cm]{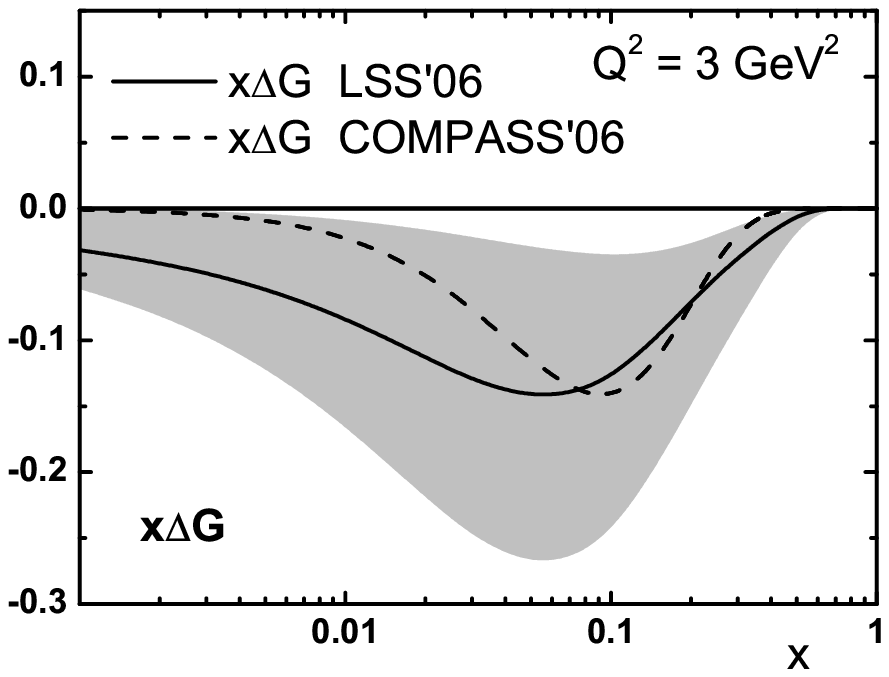}
\end{minipage}
\caption{Comparison of LSS'06 and COMPASS for positive and
negative $\Delta G$.} \label{Fig:5}
\end{figure}

%\begin{wrapfigure}{r}{0.45\columnwidth}
%\includegraphics[width=0.44\columnwidth]{DelG_LSS06.eps}
%\caption{The three different LSS'06 parametrizations of $\Delta G$
%.} \label{Fig:4}
%\end{wrapfigure}
 In Fig. \ref{Fig:4} we show
the three LSS'06 versions of $\Delta G(x)$. In  Fig. \ref{Fig:5}
we compare LSS'06 and COMPASS results for positive and negative
$\Delta G(x)$.

%\begin{figure}[h]
%\begin{center}
%\includegraphics[width=0.45\columnwidth]{PosDelG_LSSvsCOMPASS.eps}
%\caption{Comparison of LSS'06 and COMPASS for positive $\Delta
%G$.}\label{Fig:3}
%\end{center}
%\end{figure}

%\begin{figure}[h]
%\begin{center}
%\includegraphics[width=0.45\columnwidth]{NegDelG_LSSvsCOMPASS.eps}
%\caption{Comparison of LSS'06 and COMPASS for negative $\Delta
%G$.}
%\label{Fig:4}
%\end{center}
%\end{figure}
%\smallskip

%%%%%%%%%%%%%%%%%%%%%%%%%%%%%%%%%%%%%%%%%%%%%%%%%%%%%%%%%%%%%%%%%%%%%%%%%

%%%%%%%%%%%%%%%%%%%%%%%%%%%%%%%%%%%%%%%%%%%%%%%%%%%

%\begin{figure}[h]
%\begin{center}
%\includegraphics[width=0.45\columnwidth]{DelG_LSS06.eps}
%\caption{The three different LSS'06 parametrizations of $\Delta G$
%.} \label{Fig:4}
%\%end{center}
%\end{figure}
%\smallskip
%

It is seen that while the first moments are about the same, the
shapes of $\Delta G(x)$ are considerably different for the
positive case.

\begin{wrapfigure}{r}{0.4\columnwidth}
\includegraphics[width=0.40\columnwidth]{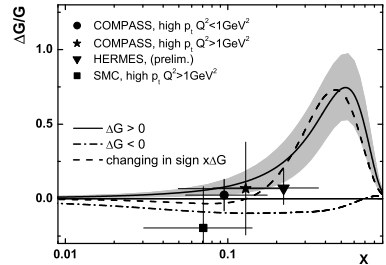}
\caption{Comparison of the COMPASS results for $\Delta G /G $ with
the three LSS'06 versions of $\Delta G $ divided by the MRST'02
$G$.} \label{Fig:6}
\end{wrapfigure}

(ii) Another approach to $\Delta G(x)$ is via $c\bar{c}$
production in SIDIS. This is based on the very reasonable
assumption that there is no intrinsic charm in the nucleon, so the
production involves $\gamma - gluon$ fusion. Ideally, to be
absolutely sure of the mechanism, one would like to detect both
charmed particles. In practice this is not possible and one relies
on single charm production at a reasonably large transverse
momentum, or on jet production. In Fig. \ref{Fig:6} we see that
the $\Delta G(x)/G(x)$ extracted by COMPASS from their data is
perfectly compatible with the three different LSS'06
parametrizations of $\Delta G(x)$ divided by the MRST'02 version
of the unpolarized gluon density $G(x)$.

%\begin{figure}[h]
%\begin{center}
%\includegraphics[width=0.45\columnwidth]{dGovG_Dubna07.eps}
%\caption{Comparison of the COMPASS results for $\Delta G /G $ with
%the three LSS'06 versions of $\Delta G $ divided by the MRST'02
%$G$.} \label{Fig:5}
%\end{center}
%\end{figure}
%\smallskip

(iii) One can also study $\Delta G(x)$ via its role in single
particle production in polarized proton-proton collisions,
especially at RHIC. We have not yet tested the LSS'06 densities by
this method.

\section{EIC and the sign of the polarized gluon density}

\begin{figure}[tbh]
\begin{minipage}[c]{7cm}
\includegraphics[width = 6cm]{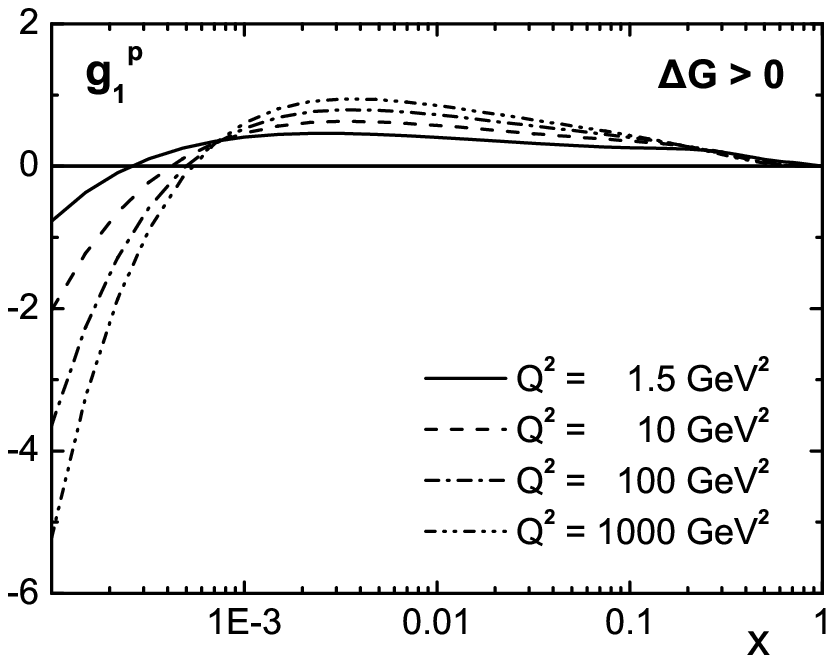}
\end{minipage}
\hspace*{0.5cm}
\begin{minipage}[c]{7cm}
%\vspace*{0.4cm}
\includegraphics[width= 6cm]{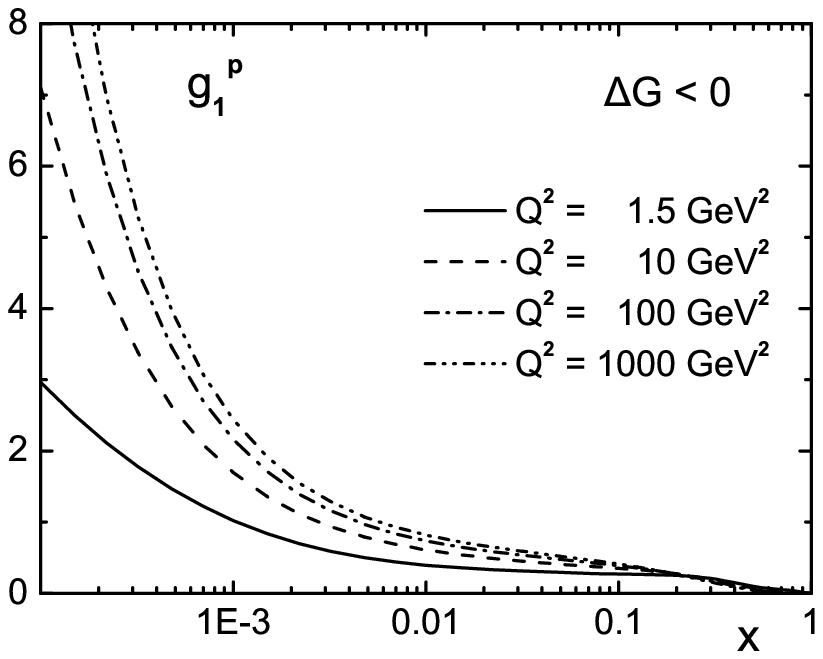}
\end{minipage}
\caption{$g_1(x,Q^2)$ at different $Q^2$ calculated using the
LSS'06  positive and negative $\Delta G$.} \label{Fig:7}
\end{figure} \vspace*{0.4cm}
It seems clear that present day data cannot distinguish between
the three scenarios for $\Delta G(x)$. A clean distinction, at
least between the positive and negative cases, would be possible
in an EIC type collider which could access values of $Q^2$ in the
region of $1000\, GeV^2$. Fig. \ref{Fig:7}  shows $g_1(x)$
 for protons, calculated using the LSS'06 positive and
negative $\Delta G(x)$, for $1\leq Q^2 \leq 1000 \,GeV^2$. There
is a dramatic difference at small $x$.

%\begin{figure}[h]
%\begin{center}
%\includegraphics[width=0.45\columnwidth]{g_1VariousQ^2_dG_pos_BW.eps}
%\caption{$g_1(x,Q^2)$ at different $Q^2$ for positive $\Delta G$.}
%\label{Fig:7}
%\end{center}
%\end{figure}
%\smallskip

%\begin{figure}[h]
%\begin{center}
%\includegraphics[width=0.45\columnwidth]{g_1VariousQ^2_dG_neg_BW.EPS}
%\caption{$g_1(x,Q^2)$ at different $Q^2$ for negative  $\Delta
%G$.}
% \label{Fig:8}
%\end{center}
%\end{figure}
%\smallskip
%%%%%%%%%%%%%%%%%%%%%%%%%%%%%%%%%%%%%%%%%%%%%%%%%%%%%%%%%%%%%%%

%%%%%%%%%%%%%%%%%%%%%%%%%%%%%%%%%%%%%%%%%%%%%%%%%%%

%\newpage

\section{Conclusions}

\begin{itemize}
\item The very accurate low $Q^2$ CLAS data significantly reduce
the errors on the HT terms.
\item Surprisingly, the increase in available data seems to
increase the freedom in the functional form of $\Delta G(x)$.
\item Positive, negative and sign-changing forms for $\Delta G(x)$
seem to be allowed with excellent $\chi^2$ values, provided HT
terms are included.
\item Measurements of $g_1(x)$ at small $x$ and large $Q^2$ at EIC
could settle the question of the sign of $\Delta G(x)$.
\end{itemize}

\end{document}